# Details Matter: noise and model structure set the relationship between cell size and cell cycle timing

**Felix Barber** [1], **Po-Yi Ho** [2], **Andrew W. Murray** [1,3,*] **and Ariel Amir** [2,*]

[1] *Department of Molecular and Cellular Biology, Harvard University, Cambridge, MA 02138, USA*

[2] *School of Engineering and Applied Sciences, Harvard University, Cambridge, MA 02138, USA*

[3] *FAS Center for Systems Biology, Harvard University, Cambridge, MA 02138, USA*

Correspondence*:
Ariel Amir, School of Engineering and Applied Sciences, Harvard University, 29 Oxford Street, Cambridge, 02138 MA, USA
arielamir@seas.harvard.edu

Andrew W. Murray
Department of Molecular and Cellular Biology, Harvard University, Cambridge, MA 02138, awm@mcb.harvard.edu

## ABSTRACT

Organisms across all domains of life regulate the size of their cells. However, the means by which this is done is poorly understood. We study two abstracted "molecular" models for size regulation: inhibitor dilution and initiator accumulation. We apply the models to two settings: bacteria like *Escherichia coli*, that grow fully before they set a division plane and divide into two equally sized cells, and cells that form a bud early in the cell division cycle, confine new growth to that bud, and divide at the connection between that bud and the mother cell, like the budding yeast *Saccharomyces cerevisiae*. In budding cells, delaying cell division until buds reach the same size as their mother leads to very weak size control, with average cell size and standard deviation of cell size increasing over time and saturating up to 100-fold higher than those values for cells that divide when the bud is still substantially smaller than its mother. In budding yeast, both inhibitor dilution or initiator accumulation models are consistent with the observation that the daughters of diploid cells add a constant volume before they divide. This "adder" behavior has also been observed in bacteria. We find that in bacteria an inhibitor dilution model produces "adder" correlations that are not robust to noise in the timing of DNA replication initiation or in the timing from initiation of DNA replication to cell division (the $C + D$ period). In contrast, in bacteria an initiator accumulation model yields robust adder correlations in the regime where noise in the timing of DNA replication initiation is much greater than noise in the $C + D$ period, as reported previously [1]. In bacteria, division into two equally sized cells does not broaden the size distribution.

# 1 KEY RESULTS AND OUTLINE

## 1.1 Key Results

- Symmetrically dividing budding cells are unable to regulate their size effectively using either an inhibitor dilution or initiator accumulation strategy. Simulations demonstrate increases in mean and standard deviation of cell sizes up to 100 fold higher than an asymmetrically dividing control for both inhibitor dilution and initiator accumulation models.
- Based on the correlation between volume at birth and division, both inhibitor dilution and initiator accumulation models can yield robust adder behavior in asymmetrically dividing, budding cells. This is consistent with observed adder behavior in budding yeast, and as such we cannot exclude either model from consideration as a viable size regulation strategy in this organism.
- It is unlikely that bacteria that display adder behavior use an inhibitor dilution strategy to regulate their cell size, since implementing such a strategy in cells that grow fully before setting their plane of division does not produce adder correlations that are robust to noise.
- An initiator accumulation model in bacteria is consistent with the experimentally observed adder behavior, provided cells grow in the regime where noise in their timing of DNA replication initiation is much greater than noise in the time from initiation of DNA replication to cell division.

## 1.2 Outline

The paper is structured as follows:

- Section 2, subsection 2.1 provides necessary background on the cell cycle in both bacteria and budding yeast, and details assumptions made throughout the text about the growth morphologies of these organisms. In subsection 2.2 we address the necessary background on size regulation in both budding yeast and bacteria. Subsection 2.3 discusses the approach of the paper. Finally, subsection 2.4 provides mathematical definitions of the two models of size regulation studied.
- Section 3 addresses cells that grow by budding, with an application to budding yeast. The growth models used for this cell type are outlined in subsection 3.1. We study this growth morphology for cells that divide asymmetrically in subsection 3.2 and for symmetrically dividing budding cells in subsection 3.3. Within these subsections we apply the inhibitor dilution (3.2.1 and 3.3.1) and initiator accumulation (3.2.2 and 3.3.2) models to the relevant cell types.
- Section 4 addresses non-budding cells, with an application to certain bacteria including *E. coli*. The cell growth model used is outlined in subsection 4.1. We consider the inhibitor dilution model in subsection 4.2 and the initiator accumulation model in subsection 4.3.
- Table 1 provides an index for the locations of model definitions used throughout the text.

# 2 INTRODUCTION

Organisms across all domains of life regulate their cell size, coupling growth and division to constrain the range of cell sizes produced. Despite this ubiquity, understanding how size control is implemented on a molecular level has remained an active area of research for several decades [2]. Two longstanding models which connect cell size with cell cycle progression are the initiator accumulation and the inhibitor dilution models [2, 3]. The initiator accumulation model involves the cyclical synthesis and degradation of an initiator protein that prompts the initiation of DNA replication. After a sufficient amount of initiator has been produced, DNA replication is initiated, and the initiator protein is subsequently degraded in its entirety so that the accumulation process must begin again from zero. In contrast, the inhibitor dilution model involves the cyclical production and dilution of a protein which inhibits initiation of DNA replication. This inhibitor protein is produced only within one part of the cell cycle, with DNA replication in the subsequent cell cycle only beginning once the inhibitor concentration has been diluted through new growth to a sufficiently low level.

## 2.1 Growth morphology and the cell cycle

In this work we apply these distinct models of size regulation to organisms that adopt two distinct modes of growth: cells that produce offspring by budding, such as the budding yeast, and non-budding cells such as the bacteria *E. coli*. We use the term non-budding to describe cells which grow fully before setting the plane of division. Results throughout will apply generally to organisms which obey the assumed growth morphologies and cell cycle structures. However, for ease of interpretation we use cell cycle structure and nomenclature appropriate to the specific examples of budding yeast and *E. coli*. These distinct modes of growth are summarized in Figure 1. In subfigure (A), newborn budding yeast cells grow during the G1 phase before passing through the cell cycle transition known as Start (the point of irreversible commitment to DNA replication and cell division) [4]. Following Start, cells replicate their DNA during S phase, and go through an additional growth phase known as G2 before entering M phase and undergoing mitosis. These phases differ in how new growth occurs in the cell, since passage through Start also represents the onset of budding, where a new bud begins to grow from the side of the existing cell. The bud grows larger throughout the S/G2/M phases before separating at the end of mitosis to form a "daughter" cell. We will use the term daughter throughout this text to describe newborn cells (i.e. buds that have just separated from the main cell) that are going through their first cell cycle. Conversely, "mother" cells must already have been through at least one independent cell cycle. During G1 the main cell body grows, but after the onset of budding, growth is predominantly given to the bud [5]. Subfigure (B) shows that in *E. coli*, newborn cells will also grow before initiating DNA replication. After initiating DNA replication, cells will wait a time $t \equiv C + D$ before dividing, where $C$ is the time for the completion of DNA replication and $D$ is the time from the completion of DNA replication to division [1]. In slow growing cells this cycle takes place between two cell division events, however, in fast growing cells the presence of multiple replication forks complicates this picture. For simplicity we only consider slow-growing *E. coli* cells in this text. In non-budding cells the division plane is set at the point of cell division. For the case of symmetrically dividing cells such as *E. coli* this is located at the midpoint of the cell so that both progeny are of roughly equal size. A key consequence of these distinct modes of growth is that for non-budding cells, growth throughout the whole cell cycle affects the size of both progeny. Further, the volume of two daughter cells at birth may be smaller than the birth volume of their parent cell due to noise. In contrast, in budding cells the size of a given cell will monotonically increase over successive generations, and the size of the main cell will not be affected by noise in budded growth. We note that throughout this work we assume that cell volume grows exponentially as a function of time, as evidenced for budding yeast and certain bacteria by highly accurate measurements of the buoyant mass of single cells [6, 7].

## 2.2 Size Regulation

Here we present the necessary background on size regulation in budding yeast and bacteria. For a broader discussion of these topics and size control in other organisms we direct readers to the following review articles: [8, 9, 10, 11]. In budding yeast, size regulation is observed in the first cell cycle of small daughter cells delaying Start relative to large daughters through a longer G1 phase [12]. Key regulators of this transition include the G1 cyclin Cln3, and its main downstream target, the transcriptional inhibitor Whi5

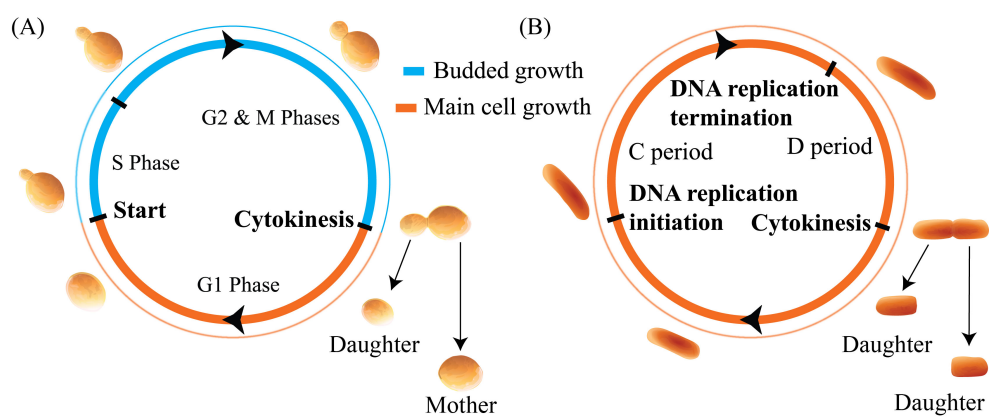

**Figure 1.** Illustration of the cell cycles and growth morphologies. (A) Budding cells (in particular the budding yeast *S. cerevisiae*). Cells grow initially, before producing a bud, setting the plane of division, and directing new growth to that bud for the remainder of the cell cycle. (B) Non-budding bacterial cells such as *E. coli*. Cells grow fully before setting the plane of division at cytokinesis. We note that in this paper we are not considering the regime of multiple replication forks in bacteria.

[13]. Whi5 is primarily localized in the nucleus during G1, where it inhibits gene expression required for DNA replication. A heterodimer composed of Cln3 and Cdk1 phosphorylates Whi5, leading to its nuclear export and activating a positive feedback loop that commits the cell to passage through Start [14, 15]. How this mechanism for cell cycle progression couples to cell size remains unclear. Recent evidence supports Whi5 being produced at a volume independent synthesis rate during the budded portion of the cell cycle [16]. This observation, combined with evidence for a volume-independent Cln3 concentration during G1 led authors to hypothesize that passage through Start couples to cell size by titrating Whi5 against Cln3. This would occur through growth-mediated dilution of nuclear Whi5, which would serve to regulate the length of the G1 phase [16, 9]. This hypothesis relies upon nuclear volume scaling with cell growth during G1, supported by constancy of the karyoplasmic ratio throughout the cell cycle [17]. Another longstanding hypothesis is that Cln3 activity may be titrated against the number of certain sites in the genome, such as the binding sites of the SBF transcription factor [13]. This hypothesis is consistent with the longstanding model that budding yeast cells grow to pass a critical size threshold regulating passage through Start [12, 18]. A third hypothesis is that the length of G1 is set instead by the integrated activity of Cln3 recorded in the Whi5 phosphorylation state, with this cumulative phosphorylation setting the timing of Whi5 nuclear export and subsequent passage through Start [19].

    Statistical correlations on single cell data now allow us to explore the connections between phenomenological models of size regulation and molecular mechanisms for cell cycle progression [5, 20, 21]. Single-cell volume measurements in diploid daughter, budding yeast cells have shown that correlations between cell size at birth and division are consistent with the phenomenological "adder" model of size regulation. Within this adder model the volume increment from birth to division is uncorrelated with the cell size at birth [5]. This "adder" behavior is phylogenetically widespread, having also been observed within a range of bacterial species such as *E. coli*, and in archaea [10, 22]. Adder correlations in diploid budding yeast cells are robust across a range of growth media, and show that passage through Start cannot follow the previously favored "size threshold" policy; a threshold in volume at Start would cause any correlation between cell size at birth and division to vanish. These correlations also highlight a potential difference in size regulation between haploid and diploid cells, since it has been noted recently

that adder behavior is not observed in haploid daughter cells [23]. Despite this, adder size correlations in diploid cells may be consistent with the molecular mechanism outlined above: in the idealized noiseless case it has been shown that a dilution model compatible with the Whi5 dilution hypothesis can reproduce these adder correlations between volume at birth and at division [5]. This is done by effectively allowing a cell to integrate a volume increment between subsequent budding events.

In bacteria the initiator accumulation model has recently been shown to allow simultaneous regulation of both cell size and the number of origins of replication, provided that DNA replication is initiated upon accumulation of a critical abundance of initiator protein per origin of replication [24, 1, 25]. Additionally, the initiator accumulation model in symmetrically dividing bacteria has been shown to yield robust adder behavior in the regime where noise in the timing of DNA replication initiation is much greater than noise in the $C + D$ period [1]. However, in bacteria there is no known evidence that definitively excludes either an initiator accumulation or inhibitor dilution model from consideration.

## 2.3 Approach

This work builds on existing phenomenological models of size regulation using two distinct "molecular" mechanisms, focusing on the effect of these size regulation strategies on the observed correlations between cell volume at birth and division. As noted earlier, the adder phenomenon has been observed within all domains of life. This observation is remarkable, given the great evolutionary distance separating organisms that have adopted this size regulation strategy. As such, for an adder size regulation strategy to be biologically relevant we expect that it should be robust to the introduction of biological noise, and we use the classification of whether this robust "adder" behavior is observed in order to characterize the models we consider. A consistent theme therefore will be the evaluation of whether these adder correlations are robust to coarse grained noise in the cell cycle. We will perform this analysis for a selection of different cell growth morphologies. Assuming a given growth morphology, we will evaluate robustness by studying adder correlations within a biologically relevant region of parameter space that was selected based on experimental observations. In this region we tested deviations from adder behavior based on the slope of a linear regression between $V_b$ and $V_d$ (volume at birth and at division). As described previously, a slope greater than $1.0$ implies poorer size control relative to the "adder" model, while a slope less than $1.0$ brings us closer to the strongest form of size control: a cell size threshold [26]. Experimental measurements showed variation in the $V_b$ vs. $V_d$ slope of roughly $1.0 \pm 0.1$ across a selection of different growth media, producing a variety of different physiological states [5]. We evaluated the $V_b$ vs. $V_d$ slopes in the sampled portions of parameter space to determine domains in which deviations from adder behavior were consistent with this experimentally observed variation. We defined the adder behavior to be robust provided that these domains were not limited to fine-tuned ranges of model parameters, i.e. they spanned broad ranges of parameter space rather than discrete pockets. This is consistent with previous studies in this area, which have demonstrated robustness by showing limiting behavior in certain regions of phase space, such as the observation that an initiator accumulation model in bacteria yields adder behavior provided that noise in initiation of DNA replication is much greater than noise in the duration of the $C + D$ period [1]. Our definition of robustness is biologically motivated by the assumption that if cells required strongly coupled noise strengths in distinct cell cycle variables to behave as adders, we would be unlikely to observe adder behavior over a range of different growth media. Figure S2 displays experimental data from previous work for diploid cells across a range of growth media, demonstrating that noise in cell division asymmetry consistently displays $CV \leq 0.3$ across a range of growth media [5]. This regime of noise strength was observed for other variables such as growth rate (see Figure S2). For noise variables relating to the expression of individual genes (i.e. $\sigma_\Delta$ or $\sigma_K$), or relating to the molecular mechanisms regulating initiation of DNA replication (i.e. $\sigma_i$ or $\sigma_s$) we do not have experimental data that could provide an accurate range of noise strengths. As such, we used an estimate based on measurements of other cell cycle noises, taking $CV \leq 0.3$ in these cases. With this methodology, and by adopting minimal assumptions regarding the cell cycle, cell growth and physiology, we decoupled the effects of variability in the size regulation machinery from those due to the physiological mode of growth of cells. Doing so allowed us to make inferences about the viability of different strategies of size regulation within different growth morphologies.

## 2.4 Model Structure

### 2.4.1 Inhibitor Dilution

The inhibitor dilution model assumes that passage through Start or initiation of DNA replication occurs upon the dilution of an inhibitor molecule $I$ below a critical concentration $c_1$. This inhibitor's expression pattern is cyclical, being synthesized exclusively in the period following the initiation of DNA replication [2]. If we consider the cases of yeast and bacteria, a generic description of the model is

$$
\begin{aligned}
V_i &= (I_b + \eta)/c_1, \\
I_d &= I_b + \tilde{\Delta}.
\end{aligned}
\tag{1}
$$

Here $V_i$ is the volume at initiation of DNA replication, $I_d$ is the inhibitor abundance at division, $I_b$ is the inhibitor abundance at birth, and $\tilde{\Delta}$ is the amount of inhibitor synthesized during that cell cycle. Here and throughout the paper, the subscripts $b$, $i$ and $d$ indicate the evaluation of a variable at entry to the current cell cycle, initiation of DNA replication (Start in budding yeast) and cell division respectively. We note that $c_1$ has the effect of setting the scale of average cell size in combination with $\langle \tilde{\Delta} \rangle$, but will not affect the correlation between volume at birth and division. We assume that at the point of division the inhibitor is distributed to both progeny according to their relative volumetric fractions. We also have introduced noise $\eta \sim \mathcal{N}(0, \sigma_s)$ in the initiation of DNA replication, with standard deviation $\sigma_s$. Note that at this stage we have not made any assumptions about the distribution of $\tilde{\Delta}$. We will consider two variants of this. The "noisy synthesis rate" synthesis model assumes the inhibitor is produced at a rate $K \sim \mathcal{N}(\langle K \rangle, \sigma_K)$ for the time $t$ between Start (initiation) and division, and defines $\tilde{\Delta} = K \times t$. In contrast the "noisy integrator" synthesis model assumes inhibitor production is uncorrelated with growth in the budded portion of the cell cycle, and defines $\tilde{\Delta} \sim \mathcal{N}(\langle \tilde{\Delta} \rangle, \sigma_\Delta)$. These distinctions affect the correlation between the amount of inhibitor produced and the noise in the timing of the cell cycle after DNA replication begins.

### 2.4.2 Initiator Accumulation

In this model a cell initiates DNA replication upon accumulation of a sufficient absolute quantity of some initiator protein $A$. This initiator is synthesized during cell growth, such that a volume increment of $\Delta V$ leads to a newly synthesized amount $\Delta A = \Delta V / c_2$ of initiator protein. Here $c_2$ is a scaling factor with units of concentration that sets the scale of the size distribution in a similar manner to $c_1$. As in the inhibitor dilution model, we assume that at cell division the initiator protein is distributed to both progeny according to their relative volumetric fractions. This process is defined by equation 2, where $A_d$ is the initiator abundance at division, $A_b$ is the initiator abundance at birth, and $A_c \sim \mathcal{N}(\langle A_c \rangle, \sigma_i)$ is the critical amount of initiator required to prompt DNA replication initiation.

$$
\begin{aligned}
A_d &= c_2(V_d - V_i) \\
V_i &= V_b - A_b/c_2 + A_c/c_2
\end{aligned}
\tag{2}
$$

Here the first line comes from the definition of initiator synthesis for a given cell volume increment, and the assumption that initiator is degraded entirely at initiation of DNA replication. The behavior of the model depends on this assumption for the particular forms of noise studied here. However, the implications of not degrading initiators following initiation have not been thoroughly investigated at this point. Note that the total new cell growth between initiation of DNA replication and cell division is $\Delta V = V_d - V_i$. The second equality comes from setting the abundance of initiator at the subsequent Start event (i.e. the sum of initiator abundance at birth $= A_b$ and new initiator produced through growth $= c_2 \times (V_i - V_b)$) equal to $A_c$. This model is a simplified case of that previously proposed for fast-growing bacteria, where we now restrict the maximum number of DNA replication forks and initiation events per cell cycle to one [1].

## 3 RESULTS: BUDDING CELLS

### 3.1 Budding Growth Models

In budding yeast it has been shown that after passage through Start, virtually all cell growth occurs in the bud (called the daughter cell in the subsequent generation) while the main cell body remains at a roughly constant volume until the subsequent G1 period [5]. This behavior may be described by defining the daughter and mother cell volumes at birth in the subsequent cell cycle from the volumes in the previous cell cycle as

$$V_b^{D,n+1} = V_d^{X,n} - V_i^{X,n},$$
$$V_b^{M,n+1} = V_i^{X,n}. \tag{3}$$

Note the use of superscripts $n$ to track distinct cell cycles, and the letters $D$ and $M$ to denote daughter cells and mother cells respectively. When we make statements independent of cell type we use the letter $X$. In the case of budding yeast we consider two means of introducing noise in the growth period between Start and cell division. Equation 4 outlines a "noisy timing" growth model, where volume at division is related to volume at initiation via a noisy exponential volume growth rate $\lambda \sim \mathcal{N}(\langle\lambda\rangle, \sigma_\lambda)$, and a noisy time from initiation to division $t \sim \mathcal{N}(\langle t\rangle, \sigma_t)$. In contrast, equation 5 describes a "noisy asymmetry" growth model where the cell adds a volume $xV_i$ between initiation of DNA replication and division, with $x \sim \mathcal{N}(\langle x\rangle, \sigma_x)$ being a dimensionless variable describing the new growth.

$$V_d = V_i \times \exp[\lambda \times t] \tag{4}$$

$$V_d = V_i \times (1 + x) \tag{5}$$

Both of these are consistent with the observation that the division asymmetry ratio $r \equiv V_b^{D,n+1}/V_b^{M,n+1}$, defined in budding cell types for any mother-daughter pair, is uncorrelated with $V_b^{X,n}$ for the parent cell in the previous generation [5]. Given this definition of $r$, in the noisy asymmetry growth model we have the exact correspondence $r = x$ for the mother-daughter pair produced after cell division, while in the noisy timing growth model we have $r = e^{\lambda t} - 1$. We note that setting cell growth noise to zero in either of these cell growth scenarios (i.e. $\sigma_t = \sigma_\lambda = 0$ or $\sigma_x = 0$ respectively) allows us to construct a mapping from a noisy synthesis rate model directly to a noisy integrator model. We may do so by defining $\sigma_\Delta \equiv \sigma_K \times t$ and $\langle\tilde{\Delta}\rangle \equiv K \times t$. However, given that we generally consider the case of non-zero cell growth noise where this mapping fails, we consider these models individually.

## 3.2 Asymmetrically Dividing Budding Cells

Here we focus on asymmetric division, where in the limit of small noise terms, the slope of a linear regression between volume at birth and at division becomes exactly 1 for the inhibitor dilution and initiator accumulation models discussed above. We now ask whether these models yield robust domains of adder behavior within the biologically relevant regimes for $\langle r \rangle$ and various noise terms, and whether they are compatible with the observed adder behavior in budding yeast.

### 3.2.1 Inhibitor Dilution

Of the inhibitor dilution models considered for this work, the one which predicted the greatest domain for adder behavior assumed a noisy integrator in inhibitor synthesis and noisy asymmetry in cell growth (see Table 1). As such, within the main text we present this variant of the inhibitor dilution model. We believe the increased domain of adder behavior predicted by this model to arise from the compression of two noise terms in growth rate $\sigma_\lambda$ and G2 timing $\sigma_t$ into one noise term in the division asymmetry $\sigma_x$, and from the decoupling of inhibitor synthesis from noise in cell growth. A numerical comparison with other variants of an inhibitor dilution model is provided in Figure S1. We note that in this model formalism with $c_1 = 1$, the parameter $\langle \tilde{\Delta} \rangle$ sets the scale for the population level volume statistics. As such, cell cycle correlations will be independent of $\langle \tilde{\Delta} \rangle$ provided the relevant noise strengths are given relative to it. Due to the equality $r = x$ for the noisy asymmetry growth model, the range of values $\langle x \rangle$ and $\sigma_x/\langle x \rangle$ used below are directly inferred from experimental data for diploid cells summarized in Figure S2.

Figure 2 shows that the inhibitor dilution model can yield robust adder behavior in asymmetrically dividing budding cells. Adder behavior is observed in the regime where noise in cell division asymmetry satisfies $\sigma_s/\langle x \rangle \leq 0.15$, and noise in inhibitor production satisfies $\sigma_\Delta/\langle \tilde{\Delta} \rangle \leq 0.2$. Experimental observations show that expression of Cln3 (a key regulator of Start) and passage through Start itself and are both noisy processes in budding yeast [27, 16]. We therefore expect that adder correlations will be robust to noise in Start over the full range of biologically relevant noise strengths, in order for an inhibitor dilution model to be compatible with the robust adder behavior observed in this organism. This is supported for daughter cells in Figure 2, where comparison of row 1 with row 2 shows that the $V_b$ vs. $V_d$ slope shows little to no change with increasing noise in passage through Start $\sigma_s/\langle \tilde{\Delta} \rangle$. This behavior was observed in all variants of the inhibitor dilution model described above, as demonstrated in Figure S1. We also see that for $\langle x \rangle = 0.5$ ($\langle r \rangle = 0.5$), adder behavior is observed over the full range of noise strengths for inhibitor production $\sigma_s/\langle \tilde{\Delta} \rangle$, and displays only a weak dependence on noise in cell division asymmetry $\sigma_x/\langle x \rangle$. This may be seen in the domain of adder behavior extending almost entirely over the range of experimentally observed $\sigma_r/\langle r \rangle$ values, and fully over the range of tested $\sigma_s/\langle \tilde{\Delta} \rangle$ values. For larger $\langle x \rangle = 0.7$ (i.e. $\langle r \rangle = 0.7$) we observe a greater dependence of the $V_b$ vs. $V_d$ slope on both $\sigma_x/\langle x \rangle$ and $\sigma_\Delta/\langle \tilde{\Delta} \rangle$, constraining the range of values in which "adder" behavior is observed. We note that inserting experimentally observed values for both $\langle r \rangle$ in $\langle x \rangle$ and $\sigma_r/\langle r \rangle$ in $\sigma_x/\langle x \rangle$ within our simulations yields a maximum slope of roughly 1.25, outside of our defined range for "adder" behavior. However, due to the error associated with inferring the ratio of two volumes based on bright field images alone it may be that these values overestimate the magnitude of $\sigma_r/\langle r \rangle$. Such an overestimation could readily shift the range of biologically observed noise to within the domain of adder behavior, and prevents us from using this study as definitive test of the relevance of this model to the case of budding yeast. Further, as mentioned in Section 2.3 we do not have adequate data to constrain the parameter $\sigma_\Delta/\langle \tilde{\Delta} \rangle$. As such, we conclude that the predictions of Figure 2 are consistent with the experimentally observed "adder" behavior of diploid daughter cells being generated by an inhibitor dilution size regulation model, pending further constraints on the biologically relevant range of parameter values. Finally, we note that similar behavior to that observed above for daughter cells was also observed for mothers (see Figure S3), although experimental evidence does not support adder behavior in mother cells (data not shown) [5].

The simulations presented in Figure 2 differ from the stochastic process described in equation 1 by imposing the additional physically realistic requirement that cell volume monotonically increases. This requirement causes deviations from analytic calculations based on equation 1 due to a subset of cells being born with a concentration of inhibitor equal to or lower than the critical concentration required to pass through Start. Following equation 1 naively without imposing this additional requirement would cause this subset of the cell population to decrease in volume prior to passing Start. To address this, we forced cell growth to monotonically increase by inserting a condition that such cells would go through Start immediately, with the same volume they had at birth. Unfortunately, this condition made calculating exact analytical expressions for the $V_b$ vs. $V_d$ linear regression slope intractable. Further, approximate calculations of linear regression slopes which ignored this additional condition deviated from the observed behavior sufficiently that they did not warrant discussion. Despite this, we can readily understand the cause of this subpopulation being born ready to pass through Start immediately. The generation of cells with a low inhibitor concentration at birth is caused by the growth of mother cells over successive generations. In the limit of small noise, mother volume at Start will increase over successive generations following the recursion relation

$$V_i^{M,n+1} = \frac{1}{1+\langle r \rangle} \times \left( V_i^{M,n} + \frac{\langle \tilde{\Delta} \rangle}{c_1} \right). \tag{6}$$

Note the re-adoption of $r$ notation here, since this statement is true for both the noisy timing and noisy asymmetry growth models in the noiseless limit. Here we have eliminated $I_b^{n+1}$ in favor of the volume at the previous Start event $V_i^n$. Equation 6 and the equivalent equation for $I_b$ evolution have fixed points $V_i^*$ and $I_b^*$ satisfying $V_i^* = I_b^*/c_1 = \langle \tilde{\Delta} \rangle/(\langle r \rangle c_1)$, indicating that within this model a mother's cell volume will eventually saturate with no new growth in G1, producing new progeny which will be born with exactly the concentration of initiator required to go through Start. Introducing noise into this system will therefore result in a finite fraction of the population being born with a concentration of inhibitor lower than that required to pass through Start, leading to the observed discrepancy between simulations which force such cells to go through Start immediately and a naive implementation of equation 1. Unless explicitly noted otherwise, all simulations presented herein imposed this additional requirement that cell volume should monotonically increase.

### 3.2.2 Initiator Accumulation

As previously noted, we have a choice of how to introduce noise into the growth morphology for yeast cells obeying the accumulation model. Here we assume a noisy asymmetry growth model, and explore the effect of varying $\langle x \rangle$, $\sigma_x/\langle x \rangle$ and $\sigma_i/\langle A_c \rangle$ in an initiator accumulation model. Again, under the assumption that $c_2 = 1$, $\langle A_c \rangle$ sets the scale for cell size distributions, but $V_b$ vs. $V_d$ correlations will be independent of $\langle A_c \rangle$ provided noise strengths are given relative to it.

Figure 3 shows that the initiator accumulation model can yield robust adder behavior in asymmetrically dividing, budding cells. Adder behavior is consistently observed provided that noise in cell division asymmetry satisfies $\sigma_x/\langle x \rangle \ll \sigma_i/\langle A_c \rangle$ and that $\sigma_i/\langle A_c \rangle$ satisfies $\sigma_i/\langle A_c \rangle \leq 0.3$. This is qualitatively similar to the observation in bacteria that adder behavior is observed provided $\sigma_i/\langle A_c \rangle \gg \lambda \sigma_t$, though in bacteria we adopt a noisy timing growth model (see [1] and Section 4.3). We observe that for daughter cells within the range of explored parameter space, adder behavior is robust to increasing noise in passage through Start $\sigma_i/\langle A_c \rangle$ over the full range of parameter values tested. For $\langle x \rangle = \langle r \rangle = 0.5$ adder behavior displays only weak dependence on increasing noise in asymmetry $\sigma_x/\langle x \rangle$. However, for $\langle x \rangle = \langle r \rangle = 0.7$ we observe greater dependence on $\sigma_x/\langle x \rangle$, with increasing $\sigma_x/\langle x \rangle$ causing the $V_b$ vs. $V_d$ slope for daughter cells to be suppressed below 1 by noise in $x$. Here deviations from adder behavior due to an increase in $\sigma_x/\langle x \rangle$ are observed in a decrease in slope (stronger size control). This contrasts with the observation for the inhibitor dilution model in Section 3.2.1 that increasing $\sigma_x/\langle x \rangle$ leads to an increase in slope (weaker size control). We note that the parameter space explored here encompasses regions of non-adder behavior. However, as described in Section 3.2.1, it is possible that experimental noise in image segmentation led to the overestimation of measured values for $\sigma_r/\langle r \rangle$. Additionally, we expect that there should be high observed noise in passage through Start and Cln3 synthesis which could indicate a large $\sigma_i/\langle A_c \rangle$ and a higher range of tolerable values for $\sigma_r/\langle r \rangle$ [27, 16]. These factors lead us to conclude that we cannot exclude the initiator accumulation model from consideration in budding yeast, and that our findings are consistent with observations on adder behavior of budding yeast daughter cells [5]. We further note that our predictions for this model appear to be consistent with the experimental observation that mother slopes are in general slightly larger than 1, as calculated based on previously published datasets [5]. This prediction contrasts with that for the dilution model in asymmetrically dividing, budding cells, where mother slopes show robust adder behavior over almost the entire area of phase space studied herein (see Figure S3).

Combining the results of this Section with those of Section 3.2.1, we conclude that despite the observation of qualitative differences in behavior, we are unable based on adder-like cell cycle correlation behavior alone to distinguish between an inhibitor dilution and initiator accumulation model within budding yeast cells. However, we also note here that the initiator accumulation model has an additional advantage of inherently producing a longer G1 phase in daughter cells than mother cells, based on their difference in cell size. This is consistent with experimental observations that mother cells have significantly shorter G1 times [27]. For the initiator accumulation model we predict that

$$t_{G1} = \frac{1}{\lambda} \left( \log \left| 1 + \frac{\langle V_i^X \rangle (1+r)}{2 \times V_b} \right| - \log |1+r| \right). \tag{7}$$

This is calculated neglecting noise in $r$ and $\lambda$ for simplicity. Here $\langle V_i^X \rangle = 2\langle A_c \rangle$ is the population average volume at Start, and $V_b$ is the volume at birth. Since for any mother-daughter pair we have that $V_b^D = r \times V_b^M$, for $r < 1$ this yields a daughter growth time in G1 which is inherently longer than that of the corresponding mother cell.

As in the case of the dilution model, we found it necessary to impose the additional requirement that cell volume should only monotonically increase. This was necessary due to a subpopulation of parent cells producing sufficient initiator before division that their progeny were born with an initiator abundance $A_b > A_c$, and would therefore otherwise decrease in volume prior to Start. We understand the generation of this subpopulation of cells as follows. In the limit of small noise, the volume at Start of mother cells followed through multiple generations follows

$$V_i^{M,n+1} = \frac{\langle A_c \rangle}{c_2} + \frac{\langle r \rangle}{1 + \langle r \rangle} V_i^{M,n}. \tag{8}$$

Note that as in Equation 6 we adopt $r$ notation here, since this statement is true for both the noisy timing and noisy asymmetry growth models in the noiseless limit. This equation has a fixed point at $V_i^* = \langle A_c \rangle (1 + \langle r \rangle)/c_2$. This indicates that over successive generations, mother cells will grow to a volume where in the budded portion of the cell cycle they produce sufficient initiator to drive them through Start immediately in the subsequent cell cycle. Introducing non-zero noise will therefore result in a finite portion of the population being born with a greater abundance of activator than needed to go through Start, necessitating the additional constraint that such cells do not decrease in volume, and go through Start immediately. The analogous behavior for the inhibitor dilution model in Section 3.2.1 leads us to believe that the generation of a subpopulation of large cells which no longer regulate their size is a consequence of the budding growth morphology, not a pathology specific to these size regulation models.

## 3.3 Symmetrically Dividing Budding Cells

Here we study symmetric division in budding cells within the previously described noisy asymmetry growth model, setting $\langle x \rangle = 1$ to yield progeny of equal size. This could also be done by setting the time spent in the budded portion of the cell cycle equal to the volume doubling time of the cell $t = t_{db} \equiv \log(2)/\langle \lambda \rangle$.

### 3.3.1 Inhibitor Dilution

For the inhibitor dilution model we assumed noisy integrator synthesis in addition to noisy asymmetry growth. Noise is inserted in passage through Start $\sigma_s$, in the amount of inhibitor produced $\sigma_\Delta$, and in the division ratio $\sigma_x$. We made similar observations to those outlined below when considering noisy integrator synthesis and a noisy timing model (data not shown).

Figure 4 displays a marked difference in behavior between (A) unphysical simulations which follow equation 1 exactly by allowing cell size to decrease if necessary, and (B) more realistic simulations in which cells born with low inhibitor abundance will pass through Start immediately at their birth volume. In (A) we see robust adder behavior for small $\sigma_s/\langle \tilde{\Delta} \rangle << 1$. However, subfigure (B) negates this result, predicting very weak size control with $V_b$ vs. $V_d$ slopes of just below 2 across the full range of noise values explored. From this observation it is clear that the majority of the population is being born such that they should go through Start immediately. This may be extracted from the numerics, but it is illustrative to understand how this inherently arises in a symmetric budding growth setting, independent of whether noise in the budded growth is taken to follow a noisy timing or noisy asymmetry rule. We do so by considering the average growth in G1 for a mother-daughter pair. Using $r$ notation for budded growth, we see that the total growth in the G1 phase for a given mother-daughter pair (LHS) obeys

$$\begin{aligned}\left\langle V_i^{M,n+1} + V_i^{D,n+1} - V_b^{M,n+1} - V_b^{D,n+1}\right\rangle &= \left\langle I_d^{X,n} - I_b^{X,n} \times (1+r)\right\rangle / c_1 \\ &= \left(\langle \tilde{\Delta} \rangle - \langle r \rangle \langle I_b^{X,n} \rangle\right)/c_1 \\ &= \frac{\langle \tilde{\Delta} \rangle}{c_1}(1 - \langle r \rangle).\end{aligned} \quad (9)$$

The first equality comes from noting that $\langle V_i^{M,n+1} + V_i^{D,n+1} \rangle = \langle I_d^{X,n} \rangle / c_1$, as well as through combining equations 1 and 3 with the definition of $r$. The second equality comes through application of equation 1 for inhibitor abundance at division. The last equality uses the result that $\langle I_b^X \rangle = \langle \tilde{\Delta} \rangle$ (derived in the Supplementary Information, equation S1), where the superscript $X$ shows that the average is taken with respect to the entire population, rather than the mother or daughter subpopulations. We see that setting $\langle r \rangle = 1$ (equivalent to $\langle x \rangle = 1$ in the noisy asymmetry model) gives on average no growth in G1 for both mother and daughter cells. Once cells are prevented from decreasing in volume during G1, we therefore expect to obtain an effective timer model of size control, where the volume at division is simply $V_d = (1 + \langle r \rangle)V_i = (1 + \langle r \rangle)V_b = 2V_b$. This would predict a $V_b$ vs. $V_d$ slope of $2$.

Pure timer based models of symmetric cell division coupled with exponential volume growth on the single cell level will lead to geometric random walks in volume, with a standard deviation that grows arbitrarily large over time [24]. Motivated by this result, we implemented repeated rounds of growth and dilution in our simulations (where each new population was seeded with cells randomly selected from the previous population) and tracked population statistics over many generations. Figure 4 (C) demonstrates that when compared to control cells ($r = 0.5$), symmetrically growing populations of cells ($r = 1$) initially dramatically increase the standard deviation in $V_b$ (volume at birth). This appears to saturate at a maximal value approximately two orders of magnitude higher than that of the asymmetric control over the extent of these simulations (simulations were run for $10^5$ volume doubling times). Figure 4 (D) shows a concomitant, lesser increase in $\langle V_b \rangle$, while (E) shows that the combination of these effects yields a $CV = \sigma(V_b)/\langle V_b \rangle$ over twofold greater than the asymmetric control. Figure 4 (F) demonstrates that prior to saturation, the increase in $\sigma(\log(V_b))$ with time is inconsistent with a pure geometric random walk in volume. Within the first 1000 doubling times in which linear increases in $\log(\sigma(\log(V_b)))$ were observed on a log-log scale, a linear regression between $\log(\sigma(\log(V_b)))$ and $\log(\text{time})$ yielded a slope of $0.16$ for the full population. A geometric random walk in cell size would yield a slope of $0.5$. The saturation of $\sigma(V_b)$ at a finite value is also inconsistent with the interpretation of a geometric random walk. These deviations from the expectation of a pure timer model of symmetric cell division may be explained by the observation that for non-zero noise values the $V_b$ vs. $V_d$ slope is slightly below the value of $2$ that would be expected for a pure timer model. We believe this to indicate the presence of some very weak size control. This may constrain the cell size distributions from growing arbitrarily broad and explain the above deviations from geometric random walk behavior. Despite this, the observation of such broad and unconstrained distributions of cell volumes in symmetrically and asymmetrically dividing, budding cells illustrates a clear problem associated with symmetric cell division in a budding growth morphology.

### 3.3.2 Initiator Accumulation

We now study the initiator accumulation model for symmetrically dividing budding cells. As in Section 3.3.1, we see in Figure S4 (A) that in an initiator accumulation model the requirement that cell volume monotonically increases leads to dramatic changes in the behavior of the $V_b$ vs. $V_d$ slope. Imposing this requirement leads to $V_b$ vs. $V_d$ slopes $\leq 2.0$, that are reduced below $2.0$ by the introduction of finite noise in both $r$ and the passage through Start. As in Section 3.3.1, we interpret this to mean that the majority of cells are being born with sufficient initiator to pass through Start immediately. We can understand this behavior by considering the abundance of activator at birth in daughter cells for the noiseless limit of $r = 1$. We see that $\langle A_b^{D,n+1} \rangle = c_2 r^2 \langle V_i^{X,n} \rangle / (1 + r)$, so that larger cells will produce progeny with a greater activator abundance at birth. For $r = 1$, we also have that $\langle V_i^D \rangle = 2\langle A_c \rangle / c_2$ (derived in the Supplementary Information, equation S2), so that $\langle A_b^D \rangle = \langle A_c \rangle$. This implies that daughter cells are on average born with sufficiently high initiator abundance that they will not grow at all during G1. This result may also be demonstrated for mother cells. As such, we expect the dominant portion of the population to follow a noisy timer model, explaining the observed $V_b$ vs. $V_d$ slope. Motivated by this result we again tested the effect of serial growth and dilution on symmetrically dividing budding cells, this time following an initiator accumulation growth policy. Results are presented in Figure S4 (B-E), and are qualitatively similar to the equivalent results for an inhibitor dilution growth policy presented in Figure 4. We therefore conclude that the inability to effectively regulate cell size is a fundamental feature of symmetrically dividing cells growing in a budding morphology, and is not a consequence of specific methods of size control.

## 4 NON-BUDDING CELLS

### 4.1 Non-Budding Cells Growth Model

In this section we consider the case of cells which do not grow by budding, the growth morphology most relevant to bacteria such as *E. coli*. Equation 10 describes this pattern of growth, where fractions $f$ and $1-f$ of volume at division $V_d$ are given to each of the two progeny respectively (labeled $D_1$ and $D_2$ for convenience), so that $f = 0.5$ gives symmetric division.

$$\begin{aligned} V_b^{D_1} &= f \times V_d \\ V_b^{D_2} &= (1-f) \times V_d \end{aligned} \quad (10)$$

We further assume that volume at division is related to volume at initiation of DNA replication by the noisy timing model described in Equation 4, and that the cell cycle period $t \equiv C + D$ is uncorrelated with the doubling time $t_{db}$ [28, 29]. In fast-growing bacteria the presence of multiple replication forks means that the division event prompted by initiation of DNA replication may not take place until later cell cycles. In this case, equation 4 would necessarily refer to the total volume at the division event prompted by that round of DNA replication initiation, but we do not consider this scenario here [1].

### 4.2 Inhibitor Dilution

In bacteria, the dilution model does not give robust adder correlations for the variety of models considered. For simplicity, we focus on the case of perfectly symmetric division in slow growing bacteria, in which the *C+D* period is shorter than the doubling time. The inhibitor dilution model has the additional requirement that initiation and division must occur alternatively. This restriction complicates analytical calculations as discussed above for budding cells. Nevertheless, simulations of two variants of the inhibitor dilution model show that the slope is sensitive to both noise in the initiation threshold and noise in the *C+D* period, implying that the inhibitor dilution models considered do not produce robust adder behavior (see Figure 5).

### 4.3 Initiator accumulation

The accumulation model can give robust adder behavior in bacteria. Since rapidly growing bacteria maintain multiple ongoing rounds of DNA replication, we consider an accumulation model in which a constant volume per origin of replication is added between replication initiation events. This model can allow for an extra round of replication initiation late in the cell cycle, through stochastically accumulating a threshold number of initiators before division. The simultaneous regulation of DNA replication and cell division allows the model to robustly recover from these stochastic events [1].

Here, we derive an analytical expression for the slope $S(V_b, V_d)$ between sizes at birth and at division, under the simplifications that cells undergo perfectly symmetric division and that cells do not undergo extra rounds of replication initiation. The slope can be written as a normalized covariance,

$$S(V_b, V_d) = \frac{\langle V_b V_d \rangle - \langle V_b \rangle \langle V_d \rangle}{\langle V_b^2 \rangle - \langle V_b \rangle^2}. \quad (11)$$

The size at birth can be written in terms of the size at the previous DNA replication initiation, $V_b = V_i \exp(\lambda(\langle t \rangle + \xi_t))/2$, where $\lambda$ is the noiseless growth rate, $\langle t \rangle \equiv C + D$, and $\xi_t$ is a Gaussian random variable with standard deviation $\sigma_t$. We can then write $S(V_b, V_d)$ in terms of $V_i$ as

$$S(V_b, V_d) = 2\left(\frac{\langle V_i V_i'\rangle - \langle V_i\rangle^2}{a^2\langle V_i^2\rangle - \langle V_i\rangle^2}\right), \quad (12)$$

where $V_i'$ is the size at the next DNA replication initiation, and $a = \langle\exp(\lambda\xi_t)\rangle$. Since $\xi_t$ is a Gaussian random variable, $a = \exp(\lambda^2\sigma_t^2/2)$. Note that factors of $\exp(\lambda\langle t\rangle)$ in the numerator and in the denominator cancel. Similarly, we can relate $V_i'$ to $V_i$ by writing $2V_i' = V_i + (\langle A_c\rangle + \xi_i)/c_2$, where $\xi_i$ is a Gaussian random variable with standard deviation $\sigma_i$. We have used the simplifications that cells undergo perfect symmetric division and that cells do not undergo extra rounds of replication initiation. Substituting into the expression for the slope, we find after simplification

$$S(V_b, V_d) = \frac{1}{a^2 + 3\left(\frac{a^2-1}{b^2}\right)}, \quad (13)$$

where $b = \sigma_i/\langle A_c\rangle$. To lowest order in small variables $\sigma_i/\langle A_c\rangle$ and $\lambda\sigma_t$, the expression becomes $S(V_b, V_d) \approx 1/(1 + 3\lambda^2\sigma_t^2/b^2)$. Hence if $\sigma_i/\langle A_c\rangle \gg \lambda\sigma_t$, the slope approaches one, as confirmed by simulations (Figure 6). For comparison with the inhibitor dilution case, Figure 6 considers the case of slow-growth ($C + D < t_{db}$). However, Eq. 13 and its derivation both hold for the fast-growth case ($C + D > t_{db}$) as well. The approximate Eq. 13 deviates from numerical results only when the fraction of cells undergoing extra initiations becomes significant at $\sigma_i/\langle A_c\rangle \gtrsim 0.3$ (Figure S5). This is a biologically unrealistic regime since experiments show that *E. coli* has $\sigma_i/\langle A_c\rangle \approx 0.1$ and $\sigma_t/\langle t_{db}\rangle \approx 0.1$ [29], where both Eq. 13 and simulations predict the observed adder behavior.

## 5 DISCUSSION

We have presented results on a selection of size regulation mechanisms applied to different growth morphologies, relevant to budding cells (in particular budding yeast) and non-budding cells such as the bacteria *E. coli*. These results are summarized both here and in Tables 2 and 3.

In asymmetrically dividing budding cells we observed that both inhibitor dilution and initiator accumulation models can give rise to robust adder behavior within specific noise regimes, and both models are consistent with the observed adder behavior in budding yeast. However, both models failed to regulate size effectively in a budding growth morphology when assuming symmetric division, predicting very weak size regulation with $V_b$ vs. $V_d$ linear regression slopes of just less than 2. This failure contrasts with the relative efficacy of these mechanisms in regulating size for a symmetrically dividing bacterial growth morphology, and illustrates the problems that an organism which divides by budding would encounter if it divided symmetrically. We hypothesize that the ineffective size regulation we have predicted for symmetric division in budding cells represents a selective pressure that contributed to the evolution of asymmetric division in budding yeast.

We found that the inhibitor dilution model in budding cells can produce robust adder behavior, provided that noise in cell division asymmetry satisfies $\sigma_x/\langle x\rangle \leq 0.15$, and that noise in inhibitor production satisfies $\sigma_\Delta/\langle\tilde{\Delta}\rangle \leq 0.2$. However, consistency with experimental observations that the adder phenomenon is robust for large asymmetry noise $\sigma_r/\langle r\rangle$ in budding yeast requires that noise be taken in the asymmetry of mother and daughter cells $\sigma_x/\langle x\rangle$, rather than in $t$ and $\lambda$ (see Figure SF1). Further, we require that noise in $r$ be independent of noise in inhibitor synthesis during that period. The restriction of our consideration to this single model may be relaxed if further measurements more accurately evaluate the strength of biological noise within relevant model parameters. Further experiments notwithstanding, in order for these noisy asymmetry and noisy integrator assumptions to be compatible with the Whi5-based inhibitor dilution molecular mechanism outlined in Section 2.2, we would require that the asymmetry ratio $r$ and the Whi5 increment $\tilde{\Delta}$ synthesized in the budded portion of the cell cycle be uncorrelated with each other. One hypothesis could be that Whi5 synthesis is integrated during the bulk of the budded phase, reaching a value $\tilde{\Delta}$ prior to division and with the remaining time in the budded phase determining the exact value of $r$. This may coincide with the observed translocation of Whi5 into the nucleus some time prior to cell division [27]. Future studies could employ rapidly maturing fluorophores to accurately determine the timing of termination of Whi5 synthesis relative to this translocation event. Further study should also be directed towards more accurately measuring the noise in total cell volume growth during the budded portion of the cell cycle. This would help clarify the compatibility of our predictions with the observed adder behavior, given the biologically measured noise strength $\sigma_r/\langle r \rangle$.

Throughout this work we have assumed that the distribution of inhibitor or initiator between mother and daughter cells at division is done according to their relative volumetric fraction, consistent with these factors being present at the same cell-body-averaged concentration in each cell. This assumption stands in apparent opposition to observations in haploids that the concentration of Whi5 in the bud nucleus is higher than in the main cell nucleus at cell division [19, 16]. Further work should investigate this discrepancy by studying the distribution of Whi5 between mother and daughter at cell division in diploid cells, since these are the relevant cell type to study adder behavior [5, 23]. Recent work has also demonstrated that adder-like behavior may be generated in a Whi5 dilution model by assuming dynamical changes in Cln3 concentration during G1, in addition to a constant abundance of Whi5 at birth [30]. Their model deviates from our own in which adder-like size regulation relies on variations in inhibitor abundance at birth, and where the inhibitor is being titrated against a factor (i.e. Cln3) that is assumed to be present at a constant concentration. A model in which Cln3 dynamics play a role in regulating cell size appears unlikely given that that single cell measurements of fluorescently labeled, stabilized mutants of Cln3 showed little variation in their concentration during G1 [16]. However, measurements of the correlation between Whi5 abundance at birth and volume at birth in wild type cells will be necessary to discriminate between the two models.

We found that the initiator accumulation model in budding cells can produce robust adder behavior, provided that noise in cell division asymmetry satisfies $\sigma_x/\langle x \rangle \ll \sigma_i/\langle A_c \rangle$ and that $\sigma_i/\langle A_c \rangle \leq 0.3$. Our predictions for this model are consistent with the robust adder behavior observed in budding yeast, diploid daughter cells, given the aforementioned uncertainty in the measurements of $\sigma_r/\langle r \rangle$ used here [5]. This model may be consistent with a different molecular mechanism for size regulation in budding yeast, whereby passage through Start is prompted by signal integration of the instantaneous kinase activity of Cln3 [19]. One possible means of aligning the initiator accumulator model with this mechanism could be to encode "initiator" abundance in the phosphorylation state of nuclear Whi5. Our implementation of the initiator accumulation model was not done with a view toward modeling this particular molecular process, so we cannot speak to the robustness of such a system.

Molecular candidates aside, our results on the robustness of cell cycle correlations for daughter cells do not allow discrimination between an inhibitor dilution or initiator accumulation model as the relevant candidate for budding yeast. However, as noted in Section 3.2.2, equation 7, we observed that the initiator accumulation model has the additional advantage of inherently producing a longer G1 phase in daughter cells than mother cells, based on their difference in cell size. In contrast, the inhibitor dilution model predicts identical G1 timing for both cell types, given the assumption maintained throughout this text that the inhibitor is distributed between mother and daughter at cell division in a manner proportional to their relative volume fractions. Note that this is consistent with the inhibitor being present at the same whole-cell-average concentration in both cell types.

In bacteria, we observed that achieving adder behavior in a symmetrically dividing inhibitor dilution model requires fine-tuning of noise in the $C + D$ period. This leads us to conclude that such a model is unlikely to be biologically relevant, in light of the robust adder correlations observed [5, 20, 21, 31, 32]. In contrast, a symmetrically dividing initiator accumulation model is robust provided that noise in the $C + D$ period of the cell cycle is smaller than noise in DNA replication initiation [1]. An initiator accumulation model also allows simultaneous regulation of the number of origins of replication [1]. Since there currently exists no definitive demonstration of a particular molecular mechanism of size control in bacteria, the finding that an inhibitor dilution model in bacteria is not robust to noise may prove useful in narrowing the range of possible molecular size control mechanisms. As such, future studies should therefore be focused around determining the molecular candidates for an initiator activation mechanism. This finding showcases the efficacy of cell cycle correlations in providing a connection between phenomenological and molecular models of size regulation.

Throughout this paper we have made the implicit assumption that size regulation can be described by the abstraction of an organism's cell cycle regulatory network to a small circuit with only a few key components. Another possibility is that size regulation is a systems level phenomena, arising from the interaction of many components in a way that cannot be mapped onto the simple circuits we have outlined. This idea is discussed in work by Robert [33]. Although such a model may be the correct description of how cells coordinate their growth with the cell division cycle, we know of no experimental evidence that convincingly demonstrates this in budding yeast or bacteria. Recent work has modeled cell size regulation in budding yeast using a data driven modeling approach [34]. By its nature such an approach must demonstrate predictive power beyond the range of the datasets used to define the model; this paper did not state or test any such predictions. In the absence of a predictive systems level description of size regulation or compelling evidence that the circuits regulating cell size were more complex, we focused on small circuit models because of their predictive power and minimal parameter fitting [5].

More broadly we observed that the consistency of inhibitor dilution or initiator accumulation models with experimental observations can depend on the assumptions surrounding the structure of those models. Determining whether these models are valid will require further experimentation on the specific molecular candidates for size regulation in a given organism, as is suggested for both budding yeast and bacteria above. We also noted that both the mode of growth and the division asymmetry can lead to significant changes in the robustness of inhibitor dilution models, as is evident when contrasting the asymmetrically dividing, budding yeast case of Figure 2 with the symmetrically dividing, bacterial case of Figure 5. These predicted differences in cell cycle correlations may prove to be a useful means of discriminating between hypotheses regarding size regulation within a specified growth morphology. This is evidenced in our work by the use of cell cycle correlations to eliminate the inhibitor dilution model of size regulation from consideration in bacteria.

### 5.1 Tables

### CONFLICT OF INTEREST STATEMENT

The authors declare that the research was conducted in the absence of any commercial or financial relationships that could be construed as a potential conflict of interest.

**Table 1.** Model Definitions Reference table

|  | Model Label | Defined in: |
|---|---|---|
| Inhibitor Dilution | Noisy synthesis rate | 2.4.1 |
|  | Noisy integrator | 2.4.1 |
| Initiator Accumulation | - | 2.4.2 |
| Growth models | Noisy Asymmetry | 3.1 |
|  | Noisy timing | 3.1 |
| Budding morphology | - | 3.1 |
| Non-Budding morphology | - | 4.1 |

**Table 2.** Size regulation summary for budding yeast mode of growth

|  | Inhibitor Dilution | Initiator Accumulation |
|---|---|---|
| Symmetric | Slope $\approx 2$ | Slope $\approx 2$ |
| Asymmetric | Robust adder | Robust adder |

**Table 3.** Size regulation summary for bacterial (non-budding) mode of growth

|  | Inhibitor Dilution | Initiator Accumulation |
|---|---|---|
| Symmetric | Not robust | Robust adder |

## AUTHOR CONTRIBUTIONS

FB performed the calculations and designed the simulations relevant to the budding yeast growth morphology. PH performed the calculations and designed the simulations relevant to the bacterial growth morphology. All authors wrote the manuscript.

## FUNDING

FB was supported during this work by the William Georgetti Trust and NIH T32 training grant number 3T32GM007598-39. PH was supported by Harvard Dean's Competitive Fund for Promising Scholarship.

## ACKNOWLEDGMENTS

FB and AA wish to thank Ilya Soifer for providing experimental data, as well as for helpful discussions and feedback. FB wishes to thank Bryan Weinstein for helpful discussions regarding implementation of the simulations. FB wishes to thank the staff of MCB graphics for assistance with the cell illustrations in Figure 1.

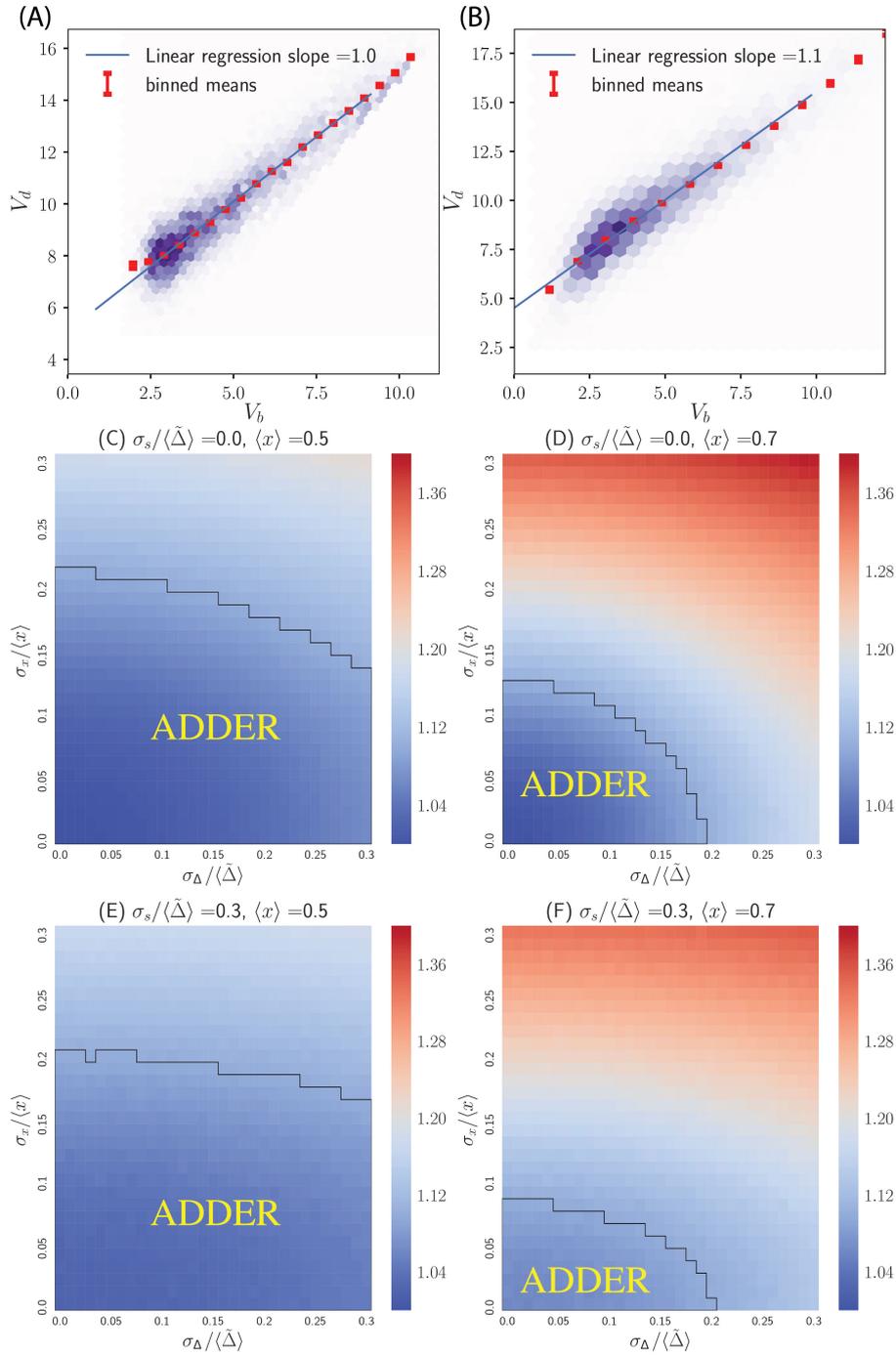

**Figure 2.** An inhibitor dilution model can yield robust adder behavior in asymmetrically dividing, budding daughter cells for biologically relevant parameter values and is consistent with experimental observations of "adder" behavior. Simulations were performed assuming noisy integrator synthesis and noisy asymmetry growth (Section 3.1). (A-B) Colored hexes denote heatmaps of volume at birth and division for daughter cells only, overlaid with a linear regression fit and red mean $V_d$ values for data points binned with respect to $V_b$. Noise values are: (A) $\sigma_x/\langle x \rangle = 0.0$, $\sigma_\Delta/\langle \tilde{\Delta} \rangle = 0.05$, $\sigma_s/\langle \tilde{\Delta} \rangle = 0.05$; (B) $\sigma_x/\langle x \rangle = 0.0$, $\sigma_\Delta/\langle \tilde{\Delta} \rangle = 0.05$, $\sigma_s/\langle \tilde{\Delta} \rangle = 0.05$. (C-F) Heat maps of linear regression slopes from fitting $V_b$ vs. $V_d$ for daughter cells only. Variation is with respect to $\sigma_\Delta/\langle \tilde{\Delta} \rangle$ and $\sigma_x/\langle x \rangle$. $\sigma_s/\langle \tilde{\Delta} \rangle$ and $\langle x \rangle$ are as labeled. Black outlines provide a guide to the eye for regions in which adder-like behavior is observed (slope $= 1.0 \pm 0.1$).

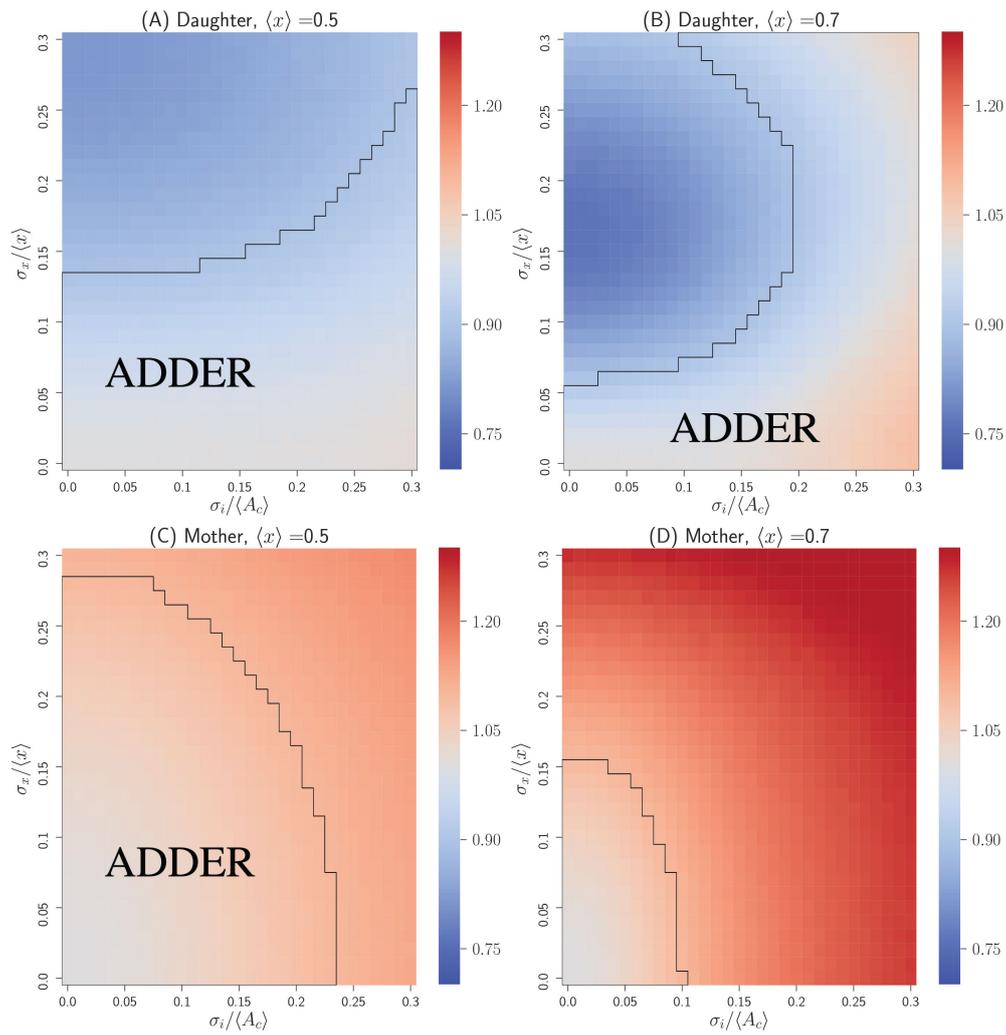

**Figure 3.** An initiator accumulation model can yield robust adder behavior for asymmetrically dividing, budding cells for biologically relevant parameter values, and is consistent with experimental observations of "adder" behavior. Heat maps are of linear regression slopes from fitting $V_b$ vs. $V_d$ for daughter or mother subpopulations in an initiator accumulation model of size regulation. (A) Daughter cells with $\langle x \rangle = 0.5$. (B) Daughter cells with $\langle x \rangle = 0.7$. (C) Mother cells with $\langle x \rangle = 0.5$. (D) Mother cells with $\langle x \rangle = 0.7$. Black outlines provide a guide to the eye for regions in which adder-like behavior is observed (slope $= 1.0 \pm 0.1$).

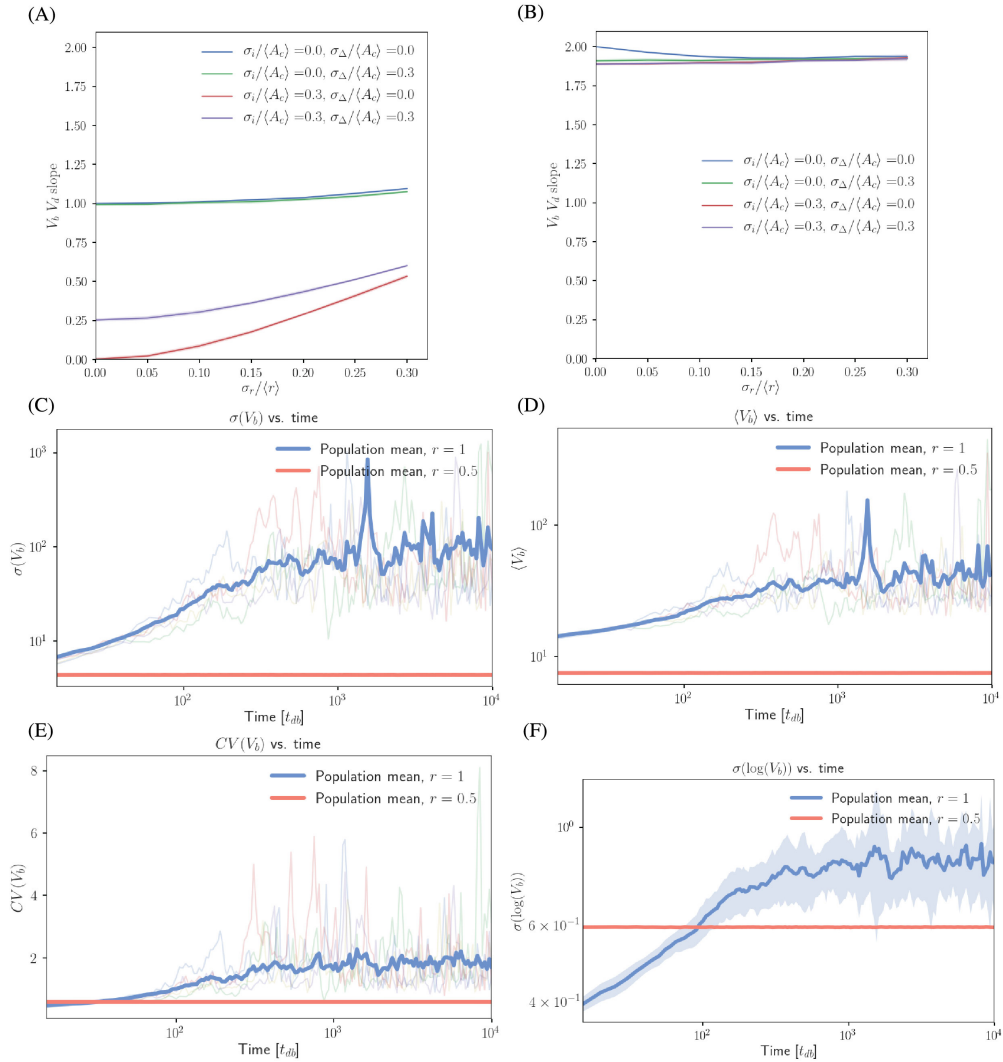

**Figure 4.** The inhibitor dilution model yields poor size regulation for symmetrically dividing, budding cells. (A-B) Linear regression slopes from fitting $V_b$ vs. $V_d$ for simulated populations of symmetrically dividing, budding cells. (A) Simulations in which cells which are born with sufficiently low inhibitor concentrations may decrease in volume before Start. (B) Simulations obtained by imposing the additional restriction that cells should pass Start with a minimum volume equal to their volume at birth. Linear regression slopes are $\leq 2$ for finite noise strengths, indicating very weak size control for symmetrically dividing, budding cells. (C-F) Population level statistics tracked for symmetrically and asymmetrically dividing, budding cells growing with an inhibitor dilution model. Sequential growth dilution steps allowed the population to be grown for $2.5 \times 10^5$ doubling times. (C) $\sigma(V_b)$ vs. time (in $t_{db}$) shows the increase in $\sigma(V_b)$ of approximately two orders of magnitude relative to the asymmetrically dividing control, with the mean traces for the 20 replicate simulated populations shown for both symmetric ($\langle x \rangle = 1$) and asymmetric ($\langle x \rangle = 0.5$) growth. Faded lines show 5 randomly selected single traces. (D) The same for $\langle V_b \rangle$ vs. time (in $t_{db}$). (E) The same for $CV(V_b)$ vs. time (in $t_{db}$). (F) $\sigma(\log(V_b))$ vs. time (in $t_{db}$) shows that the increase in standard deviation is below that expected from a pure geometric random walk in volume, with the mean for 20 repeats shown in bold. Noise was introduced in $r$, passage through Start and inhibitor production with $\sigma_x/\langle x \rangle = 0.2$, $\sigma_s/\langle \tilde{\Delta} \rangle = 0.0$, $\sigma_\Delta/\langle \tilde{\Delta} \rangle = 0.0$. Shaded error represents standard deviation of the 20 repeats tracked per condition. The saturation of the increase in average volume and standard deviation are inconsistent with a geometric random walk, but demonstrate dramatic spreading of the cell size distributions consistent with very weak size control.

*Barber et al.* Details Matter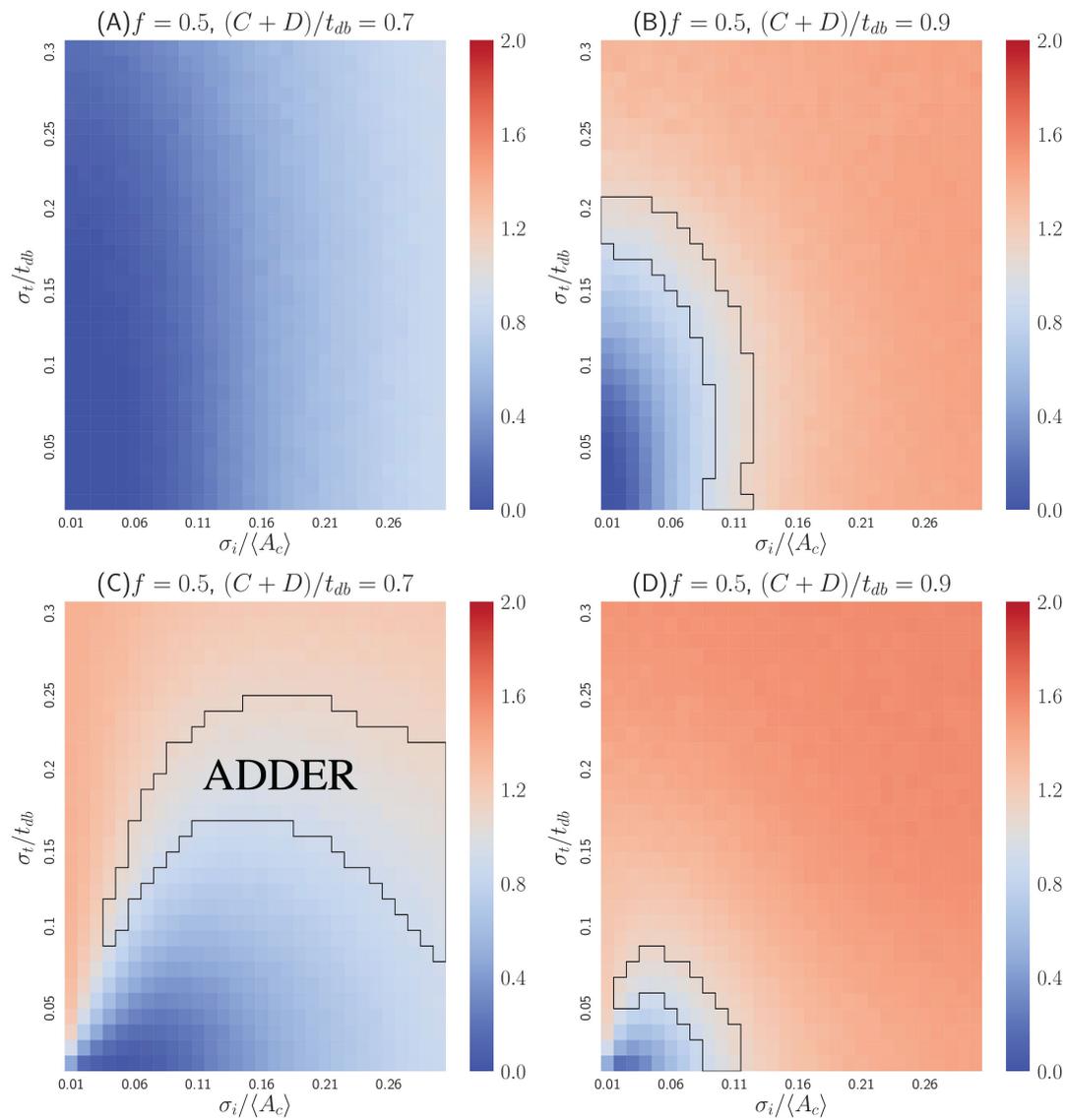

**Figure 5.** The inhibitor dilution model is not robust for the symmetrically dividing bacterial mode of growth. Heat maps of linear regression slopes from fitting $V_b$ vs. $V_d$ for symmetrically dividing bacterial cells. The models simulated are variants of an inhibitor dilution model in which the amount of inhibitor synthesized is (A-B) uncorrelated with the time spent post DNA replication initiation (noisy integrator), or (C-D) equal to a constant rate multiplied by the time spent post DNA replication initiation (noisy synthesis rate, with $\sigma_K = 0$). Black outlines provide a guide to the eye for regions in which adder-like behavior is observed (slope $= 1.0 \pm 0.1$). Adder behavior is seen to be sensitive to noise strength. This indicates that the dilution model is unlikely to be implemented as a means of size regulation in symmetrically dividing bacteria which display adder behavior.

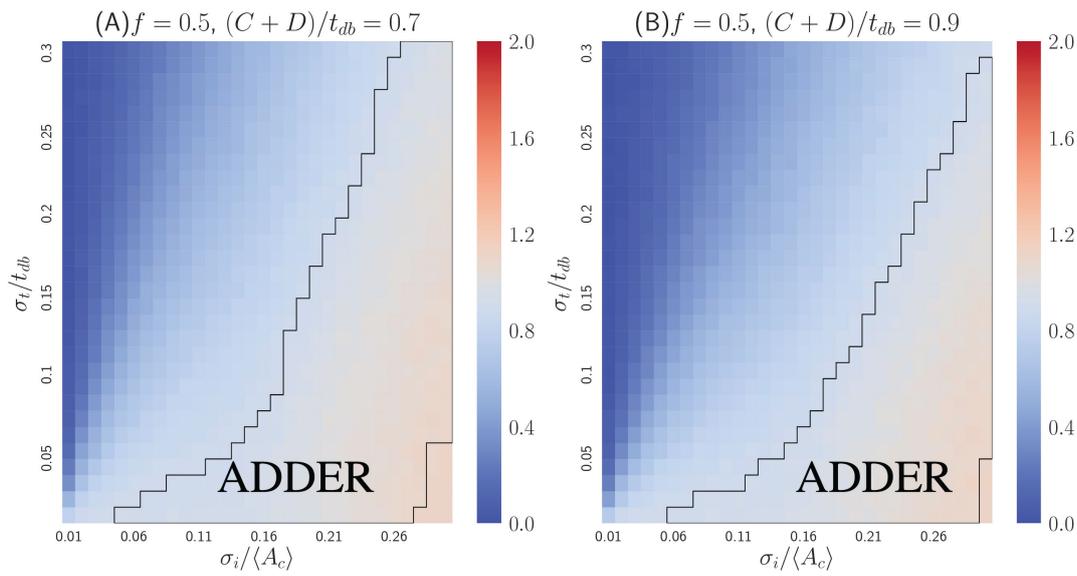

**Figure 6.** The initiator accumulation model is robust for the symmetrically dividing bacterial mode of growth, provided that $\sigma_i/\langle A_c \rangle \gg \sigma_t/t_{db}$. Heat maps of linear regression slopes from fitting $V_b$ vs. $V_d$ for symmetrically dividing bacterial cells. The model simulated is the initiator accumulation model. Black outlines provide a guide to the eye for regions in which adder-like behavior is observed (slope $= 1.0 \pm 0.1$). In the limit of $\sigma_i/\langle A_c \rangle \gg \sigma_t/t_{db}$ the observed $V_b$ vs. $V_d$ slopes approach 1, consistent with experimental observations.